\documentclass[pra,aps,twocolumn,notitlepage,superscriptaddress,showpacs,nofootinbib]{revtex4-2}
\usepackage{enumerate,appendix}
\usepackage{amsmath, amsthm, amssymb,commath,braket}
\usepackage{color,calc,graphicx}
\usepackage[usenames,dvipsnames,svgnames,table,cmyk,hyperref]{xcolor}
\usepackage[colorlinks]{hyperref}
\hypersetup{
	colorlinks = true,
	urlcolor = {blue},
	citecolor = {blue},
	linkcolor= {blue}
}

\usepackage{graphicx}
\usepackage{amsmath}
\usepackage{latexsym}
\usepackage{bbm}

\newcommand\tr{{\rm Tr }}
\newcommand\mM{{\mathcal M }}
\newcommand\mN{{\mathcal N }}

\usepackage[charter,cal=cmcal,sfscaled=false]{mathdesign}
\usepackage{booktabs}
\usepackage{multirow}
\usepackage{subfigure}
\usepackage{dcolumn}
\usepackage{mathrsfs}

\begin{document}
\title{Measurement uncertainty relation for three observables}


\author{Sixia Yu}
\affiliation{Hefei National Laboratory for Physical Sciences at Microscale and Department of Modern Physics, University of Science and Technology of China, Hefei, Anhui 230026, China}

\author{Ya-Li Mao}

\affiliation{Shenzhen Institute for Quantum Science and Engineering and Department of Physics, Southern University of Science and Technology, Shenzhen, 518055, China}

\author{Chang Niu}
\affiliation{Hefei National Laboratory for Physical Sciences at Microscale and Department of Modern Physics, University of Science and Technology of China, Hefei, Anhui 230026, China}

\author{Hu Chen}
\affiliation{Shenzhen Institute for Quantum Science and Engineering and Department of Physics, Southern University of Science and Technology, Shenzhen, 518055, China}
\affiliation{Guangdong Provincial Key Laboratory of Quantum Science and Engineering, Southern University of Science and Technology, Shenzhen, 518055, China}

\author{Zheng-Da Li}
\affiliation{Shenzhen Institute for Quantum Science and Engineering and Department of Physics, Southern University of Science and Technology, Shenzhen, 518055, China}
\affiliation{Guangdong Provincial Key Laboratory of Quantum Science and Engineering, Southern University of Science and Technology, Shenzhen, 518055, China}

\author{Jingyun Fan}
\affiliation{Shenzhen Institute for Quantum Science and Engineering and Department of Physics, Southern University of Science and Technology, Shenzhen, 518055, China}
\affiliation{Guangdong Provincial Key Laboratory of Quantum Science and Engineering, Southern University of Science and Technology, Shenzhen, 518055, China}
\affiliation{Center for Advanced Light Source, Southern University of Science and Technology, Shenzhen, 518055, China}

\begin{abstract}
 In this work we establish rigorously a measurement uncertainty relation (MUR) for three unbiased qubit observables, which was previously shown to hold true under some presumptions. The triplet MUR states that the uncertainty, which is quantified by the total statistic distance between the target observables and the jointly implemented observables,  is lower bounded by an incompatibility measure that reflects the joint measurement conditions. We derive a necessary and sufficient condition for the triplet MUR to be saturated and the corresponding optimal measurement.  To facilitate experimental tests of MURs we propose a straightforward implementation of the optimal joint measurements. The exact values of incompatibility measure are analytically calculated for some symmetric triplets when the corresponding triplet MURs are not saturated.
We anticipate that our work may enrich the understanding of quantum incompatibility in terms of MURs and inspire further applications in quantum information science. This work presents a complete theory relevant to a parallel work [Y.-L. Mao, {\it et al.}, Testing Heisenberg's measurement uncertainty relation of three observables, arXiv:2211.09389] on experimental tests.

\end{abstract}

\maketitle 

\section{Introduction}
One of the most distinguishing features of quantum theory, from a modern point of view, is the  incompatibility~\cite{Heinosaari_2016, Guhne2021} of quantum measurements whose outcomes cannot be read out simultaneously. From a practical point of view the quantum incompatibility, just like entanglement \cite{GUHNE20091,RevModPhys.81.865}, can be established as a non-classical resource and  has found numerous applications in quantum informations science including Bell's nonlocality, quantum steering, quantum contextuality, and quantum state discrimination. From a fundamental point of view, the quantum incompatibility quantifies Bohr's complementarity and lies at the heart of all kinds of Heisenberg's uncertainty relations~\cite{Heisenberg1927Physik}. 

By employing different measures for uncertainty (or errors or disturbance) such as mean square errors, trace distance, or entropies, various kinds of uncertainty relations have been established~\cite{BUSCH20061,BUSCH2007155,RevModPhys.82.1155,Giovannetti2011nphoton,LIANG20111,RevModPhys.84.1655,Busch2014RMP,Coles2017RMP,RevModPhys.91.025001,RevModPhys.94.025008}. These uncertainty relations can be classified into two different kinds, namely, 
the preparation uncertainty relations (PUR, also known as the Heisenberg-Robertson uncertainty relation) ~\cite{Kennard1927Physik,Robertson1929PhysRev} and the measurement uncertainty relations (MUR) ~\cite{RevModPhys.42.358,Ozawa2003PRA,Werner2004,Ozawa2004AOP,Ozawa2004PLA,Hofmann2003PRA,Hall2004PRA,Lund2010NJP,Fujikawa2012PRA,Branciard2013PNAS,PhysRevLett.110.010404,Busch2013PRL,Busch2014PRA,Yu2014arXiv,Erhart2012Nphysics,Rozema2012PRL,Sulyok2013PRA,Baek2013SP,PhysRevLett.110.220402,Ringbauer2014PRL,Kaneda2014PRL,PhysRevLett.114.070402,Ma2016PRL,Zhou2016SA,Xiong2017NJP,Bullock_2018,Mao2019PRL,yunger2019entropic}. While the PURs prohibit us from preparing quantum states with definite values for incompatible observables, the MURs capture the essence of quantum incompatibility, namely, quantum measurements may disturb each other, which was the main concern in the original  Heisenberg's gedanken experiment of microscope~\cite{Busch2013PRL, Busch2014PRA}. Recent years have witnessed a number of MURs of two observables being conceived and verified in experiments ~\cite{Ozawa2003PRA,Ozawa2004AOP,Ozawa2004PLA,Hofmann2003PRA,Hall2004PRA,Lund2010NJP,Fujikawa2012PRA,Branciard2013PNAS,PhysRevLett.110.010404,Busch2013PRL,Busch2014PRA,Yu2014arXiv,Erhart2012Nphysics,Rozema2012PRL,Sulyok2013PRA,Baek2013SP,PhysRevLett.110.220402,Ringbauer2014PRL,Kaneda2014PRL,PhysRevLett.114.070402,Ma2016PRL,Zhou2016SA,Xiong2017NJP,Bullock_2018,Mao2019PRL,yungerExp}.

The quantum incompatibility can be quantified most naturally in the framework of joint measurement. For example,  we can approximately measure two incompatible observables by performing a pair of compatible, i.e., jointly measurable, quantum observables by introducing some errors. The resulting MUR indicates that the total error, which may be quantified in, e.g., a worst case senario to characterize the performance of the measuring device in the spirit of Bush, Lahti,  and Werner (BLW)~\cite{Busch2014PRA}, is lower bounded by some measure of incompatibility. The optimal errors, however, are in generally achieved on different states.

An  operational and meaningful lower bound for the total error was proposed independently in Refs.\cite{Ma2016PRL, math2018}, via the distance, e.g., relative entropy or statistic distance, from the given pair of incompatible measurements to the set of all the jointly measurable pairs, calculated on the same state.  For relative entropy approach the lower bounds as well as optimal measurements are in general not analytically tractable. Moreover a relation of the lower bound to the joint measurability condition is missing. For the statistic distance measure of total error, an elegant form of MUR is proposed with a presumption that approximate joint measurements are restricted to the unbiased triplets  \cite{Qin2019PRA,EP2022}. Thus a MUR for a triplet of observables is still missing and how to find the optimal measurements to attain the lower bound and their experimental implementations remain to be outstanding problems.

In this paper we at first re-establish the MUR for triplet of unbiased qubit observables by considering the most general jointly measurable triplet as approximations. And then we provide a necessary and sufficient condition of attainability as well as the optimal measurement. Third, we  propose a  straightforward implementation of the optimal joint measurement by measuring randomly  four ideal qubit observables. Lastly, we derive analytically the incompatibility measure for two symmetric triplets after showing that the optimal jointly measurable triplet shares the same graded symmetries with the original triplet of incompatible observables.

\section{Triplet MUR and an implementation}
For a qubit the most general measurement or observable with two outcomes is represented by positive operator valued measure (POVM) $\{N_\pm\}$ with
\begin{equation}
 N_{\pm}=\frac{  {I}\pm(x+{\vec{n}}\cdot\vec{\sigma})}{2}:= N_{\pm}(x,\vec n),
\end{equation}
where we have denoted explicitly its dependence on the biasedness $x$ and Bloch vector $\vec n$ satisfying $|x|+|\vec n|\le1$.  The observables with vanishing biasedness, i.e. $x=0$, are referred to as unbiased.
In general, some POVMs are called jointly measurable if there exists a parent POVM with multiple outcomes such that each observable in the given set arises as a marginal measurement or equivalently from a post-measurement processing~\cite{Toigo2009}. The exact joint measurement conditions are known in a few special cases~\cite{PhysRevD.33.2253,Yu2010,Yu2013arXiv,PhysRevA2016} and in general the problem can be cast into semidefinite programmings~\cite{PhysRevLett2009,Schneeloch2013PRA}. 

In particular,  a set $\mN=\{N_j\}_{j=1}^3$ of three general two-outcome qubit observables $N_j=\{ N_{\pm|j}=N_{\pm}(x_j,\vec n_j)\}\}$ is jointly measurable iff there exists a parent measurement
 $\{   M({\mu_1,\mu_2,\mu_3})\}$ with 8 outcomes, labelled with binary vector $\mu=(\mu_1,\mu_2,\mu_3)$ for $\mu_j=\pm1$, such that the given three general observables arise as marginals, i.e., for each $j=1,2,3$ it holds
\begin{equation} N_{\mu_j|j}=\prod_{k\not=j}\sum_{\mu_k=\pm} M({\mu_1,\mu_2,\mu_3}).
\end{equation}
Equivalently, $\mN$ is jointly measurable iff there exists a parent measurement $\{   M_\omega\}$ together with a set of post-measurement processing $\{p_j(\pm|\omega)\}$, i.e., a set of probability distributions for each $j$ and outcome $\omega$, such that
\begin{equation} N_{\pm|j}=\sum_\omega p_j(\pm|\omega)   M_\omega.
\end{equation}
The necessary and sufficient conditions for a triplet of unbiased qubit observables to be compatible is explicitly given by~\cite{Yu2013arXiv}
\begin{equation}
	\sum_{k=0}^{3}|\vec{q}_k-\vec{q}_f|\le 4,
	\label{FMC}
	\end{equation}
where $$\vec q_k=\sum_{j=1}^3\gamma_{jk}\vec n_j,\quad \gamma_{jk}=(-1)^{j\lfloor\frac k2\rfloor+k\lfloor\frac j2\rfloor}$$
with $\vec{q}_f$ being the Fermat-Torricelli (FT) point of $\{\vec{q}_k\}^3_{k=0}$, the vector that minimize the distance sum as given in the left-hand-side of inequality Eq.(\ref{FMC}). We note that $\{\vec q_k\}_{k=0}^3=\{\mu\vec n_1+\nu\vec n_2+\mu\nu\vec n_3\mid\mu,\nu=\pm\}$.

{\bf Lemma } If a triplet  $\{   N_\pm(x_j,\vec n_j)\}_{j=1}^3$ is jointly measurable, then the corresponding unbiased triplet $\{   N_\pm(0,\vec n_j)\}_{j=1}^3$ is also jointly measurable.

{\bf Proof } As the triplet $\{   N_\pm(x_j,\vec n_j)\}$ is jointly measurable, there exists a joint measurement 
\begin{eqnarray*}
 8   M({\mu_1,\mu_2,\mu_3})=1+\sum_{j=1}^3\mu_j(x_j+\vec n_j\cdot\vec\sigma)\\+\sum_{j>k}\mu_j\mu_k(z_{jk}+\vec z_{jk}\cdot\vec\sigma)-\mu_1\mu_2\mu_3(z+\vec z\cdot\vec\sigma)\hskip-1.2cm
\end{eqnarray*}
for some real $\{z,z_{jk}\}$ and vectors $\{\vec z_{jk},\vec z\}$ $(j>k=1,2,3)$ with the given triplet as marginals. The positivity requirements $   M({\mu_1,\mu_2,\mu_3})\ge0$, i.e., demand that
\begin{eqnarray*}
1+\sum_{j=1}^3\mu_jx_j+\sum_{j>k}\mu_j\mu_kz_{jk}-\mu_1\mu_2\mu_3z\\\ge \left|\textstyle\sum_{j=1}^3\mu_j\vec n_j+\sum_{j>k}\mu_j\mu_k\vec z_{jk}-\mu_1\mu_2\mu_3\vec z\right|.\hskip-1.2cm
\end{eqnarray*}
By summing up over all $\mu_j=\pm$ we obtain
\begin{eqnarray*}
8&\ge&\sum_{\mu_1,\mu_2,\mu_3}\left|\textstyle\sum_{j=1}^3\mu_j\vec n_j+\sum_{j>k}\mu_j\mu_k\vec z_{jk}-\mu_1\mu_2\mu_3\vec z\right|\\&=&\frac12\sum_{\mu_1,\mu_2,\mu_3,\pm}\left|\textstyle\sum_{j=1}^3\mu_j\vec n_j\pm\sum_{j>k}\mu_j\mu_k\vec z_{jk}-\mu_1\mu_2\mu_3\vec z\right|\\
&\ge&\sum_{\mu_1,\mu_2,\mu_3}\left|\textstyle\sum_{j=1}^3\mu_j\vec n_j-\mu_1\mu_2\mu_3\vec z\right|\\&=&2\sum_{\mu_1\mu_2=\mu_3}\left|\textstyle\sum_{j=1}^3\mu_j\vec n_j-\vec z\right|\ge2\sum_{\mu_1\mu_2=\mu_3}\left|\textstyle\sum_{j=1}^3\mu_j\vec n_j-\vec q_f\right|\\
\end{eqnarray*}
where the first equality is due to an average over $\vec \mu$ and $-\vec \mu$ while the first inequality is because of triangle inequality and the last inequality comes from the definition of FT point $\vec q_f$ of four vectors
$\{\mu_1\vec n_1+\mu_2\vec n_2+\mu_1\mu_2\vec n_3\}_{\mu_1,\mu_2=\pm}=\{\vec q_k\}_{k=0}^3$. 
This means that the unbiased triplet $\{   N_\pm(0,\vec n_j)\}$ is also jointly measurable. \hfill $\square$

Let $\mM=\{M_j\}_{j=1}^3$ with $M_j=\{ M_{\pm|j}:=   N_{\pm}(0,\vec m_j)\}$ be a triplet of unbiased qubit observables. A jointly measurable triplet of general observables $\mN=\{   N_{\pm|j}=   N_{\pm}(x_j,\vec n_j)\}$ is performed and the total uncertainty measured by the statistics distance defines the incompatibility of the triplet 
\begin{equation}
	\Delta_\mM:=\min_{\mN}\max_\rho\sum_{i=1}^3d_{\rho}( M_i;  N_i)\end{equation}
	where
	\begin{equation} d_{\rho}( M_i;  N_i)=2\sum_\pm\left|\tr\rho M_{\pm|j}-\tr\rho N_{\pm|j}\right|
\end{equation}
An elegant lower bound of $\Delta_\mM$ was proposed in~\cite{Qin2019PRA} which is intimately related to the joint measurability condition. However, in deriving their result a strong presumption that the optimal measurements are unbiased was introduced. We strengthen this measurement uncertainty relation by considering the most general form of jointly measurable triplet by proving that  the optimal measurement is actually unbiased.

{\bf Theorem 1 } (Triplet MUR) For an unbiased triplet  $\mM=\{M_{j}\}_{j=1}^3$,
by performing the most general measurements $\{N_j\}_{j=1}^3$ that are jointly measurable, it holds MUR
\begin{equation}\label{MUR}
	\Delta_\mM\ge \frac{1}{2}\sum_{k=0}^{3}|\vec{p}_k-\vec{p}_f|-2:=2\delta
\end{equation}
where $\{\vec p_k=\sum_j\gamma_{jk}\vec m_j\}$ with $\vec p_f$ being its FT point. The lower bound is saturated if and only if
\begin{equation}
\delta\le	\min_k|\vec p_k-\vec p_f|. 
\end{equation}
If the condition is met, the optimal set of jointly measurable triplet reads
\begin{equation}\label{optm}
\vec n_{j}=\vec m_{j}+\frac\delta4\sum_{k=0}^3\gamma_{jk}\frac{\vec p_f-\vec p_{k}}{|\vec p_f-\vec p_{k}|}\quad (k=1,2,3).
\end{equation}

{\bf Proof } To lower bound the incompatibility we calculate
\begin{widetext}
\begin{subequations}\label{unn}
\begin{eqnarray}
\frac12\Delta_\mM&=&\min_{\mN}\max_\rho \textstyle\sum_{j=1}^3\sum_\pm\left|\tr\rho(M_{\pm|j}-N_{\pm|j})\right|\\
&=&{\min_{\{x_j,\vec n_j\}}\max_{\vec r}}\textstyle\sum_{j=1}^3|\vec r\cdot(\vec m_j-\vec n_j)-x_j|\\
&=&\min_{\{x_j,\vec n_j\}}\max_{\vec r}\max_{\mu_1,\mu_2,\mu_3=\pm1}\textstyle\sum_{j=1}^3\mu_j\left(\vec r\cdot(\vec m_j-\vec n_j)-x_j\right)\label{abs}\\
&=&\min_{\{x_j,\vec n_j\}}\max_{\mu_1,\mu_2,\mu_3=\pm1}\max_{\vec r}\vec r\cdot\left(\textstyle\sum_{j}\mu_j\vec m_j-\sum_{j}\mu_j\vec n_j\right)-\textstyle\sum_{j}\mu_jx_j\\
&=&\min_{\{x_j,\vec n_j\}}\max_{\mu,\nu=\pm}\left|\textstyle \mu\vec m_1+\nu\vec m_2+\mu\nu\vec m_3-(\mu\vec n_1+\nu\vec n_2+\mu\nu\vec n_3)\right|+\left|\mu x_1+\nu x_2+\mu\nu x_3\right|\label{mn}\\
\mbox{condition 0}\quad&\ge&\min_{\{\vec n_j\}}\max_{k=0,1,2,3}|\vec p_{k}-\vec q_{k}|\label{un}\label{c0}\\
\mbox{condition 1}\quad&\ge&\min_{\{\vec n_j\}}\frac14\textstyle\sum_{k=0}^3|\vec p_{k}-\vec q_{k}|\\
\mbox{condition 2}\quad&\ge&\min_{\{\vec n_j\}}\frac14\textstyle\sum_{k=0}^3\big||\vec p_{k}-\vec q_f|-|\vec q_{k}-\vec q_f|\big|\\
\mbox{condition 3}\quad&\ge&\min_{\{\vec n_j\}}\frac14\textstyle\sum_{k=0}^3\big(|\vec p_{k}-\vec q_f|-|\vec q_{k}-\vec q_f|\big)\\
\mbox{condition 4}\quad&\ge&\frac14{\textstyle\sum_{k=0}^3|\vec p_{k}-\vec p_f|}-\frac14\max_{\{\vec n_j\}}\textstyle\sum_{k=0}^3|\vec q_{k}-\vec q_f|\\
\mbox{condition 5}\quad&\ge& \frac14{\textstyle\sum_{k=0}^3|\vec p_{k}-\vec p_f|}-1=\delta
\end{eqnarray}	
\end{subequations}
\end{widetext}
Here we have used the fact that the absolute value can be rewritten as $|A|=\max_{\pm}\{\pm A\}$ in deriving Eq.(\ref{abs}) and the fact that
$\sum_j\mu_j\vec m_j=\mu_1\mu_2\mu_3\sum_{(i,j,k)\ \rm cyclic}\mu_i\mu_j\vec m_k$ and redefine $\mu=\mu_2\mu_3,\nu=\mu_1\mu_3$ in deriving Eq.(\ref{mn}).  In the first inequality yielding condition 0 we have  used Lemma to take $x_j=0$ while keeping the resulting unbiased set still jointly measurable with a no larger total uncertainty.  In derive condition 2 we have taken $\vec q_f$ to be the FT point of $\{\vec q_k\}$. In condition 4 we have used the property of the FT point $\vec p_f$ for $\{\vec p_k\}$ while the joint measurement condition is used in deriving condition 5.

In order to saturate the MUR in Eq.(\ref{MUR}) we have only to let all the inequalities become equalities in the above derivation. It is straightforward to see that in order to saturate Eq.(\ref{c0}), i.e., the condition 0, the optimal measurement has to be unbiased. For the other inequalities we have, respectively,
\begin{enumerate}
\item $|\vec p_{k}-\vec q_{k}|$ is independent of $k$
\item $\vec p_{k},\vec q_{k},\vec q_f$ are linearly dependent for each $k$ 
\item $|\vec p_{k}-\vec q_f|\ge |\vec q_{k}-\vec q_f|$ for each $k$
\item $\vec q_f$ coincides with $\vec p_f$ 
\item $\sum_{k}|\vec q_{k}-\vec q_f|=4$.
\end{enumerate}

From conditions 2 and 4 it follows that we can have linear expansions
$\vec q_{k}=(1-\alpha_{k})\vec p_{k}+\alpha_{k} \vec p_f$
with some real coefficient $\alpha_{k}$ (arbitrary for now) for each $k=0,1,2,3$. As a result of condition 1 we have
$$\delta=|\vec p_{k}-\vec q_{k}|=|\alpha_{k}|\cdot |\vec p_{k}-\vec p_f|$$
is independent of $k$.
Therefore we can assume $|\vec p_{k}-\vec p_f|>0$ for all $k$ because otherwise $|\vec p_{k}-\vec q_{k}|$ would become zero for all $k$ (condition 1) so that the triplet $\mM$ is also joint measurable, a contradiction. Therefore from conditions 3 and 4, i.e.,
$$|\vec q_{k}-\vec q_f|=|\vec q_{k}-\vec p_f|=|1-\alpha_{k}|\cdot |\vec p_{k}-\vec p_f|\le |\vec p_{k}-\vec p_f|$$
we obtain
$|1-\alpha_{k}|\le 1$ so that we have $0\le\alpha_{k}\le 2$. Thus
$$\alpha_{k}=\frac{\delta}{|\vec p_{k}-\vec p_f|}.$$
From condition 5 it follows
\begin{eqnarray*}
4&=&\sum_{k}|\vec q_{k}-\vec q_f|=\sum_{k}|\vec q_{k}-\vec p_f|\\
&=&\sum_{k}|1-\alpha_{k}|\cdot |\vec p_{k}-\vec p_f|\\
&=&\sum_{k}\Big||\vec p_{k}-\vec p_f|-\delta\Big|\\
&\ge& \sum_{k}\big(|\vec p_{k}-\vec p_f|-\delta\big)\\
&=&4(\delta+1)-4\delta=4
\end{eqnarray*}
That is, the inequality is in fact an equality, meaning that  for all $k=0,1,2,3$ it holds
$$\frac14\sum_k|\vec p_k-\vec p_f|-1=\delta\le|\vec p_{k}-\vec p_f|.$$

Sufficiency. Suppose the condition is satisfied we have $|\vec p_{k}-\vec p_f|>0$ for all $k$ and need to show that the unbiased triplet defined in Eq.(\ref{optm}) is jointly measurable and saturate the MUR. By construction we have
$$\vec q_{k}=\sum_{j=1}^3\gamma_{jk}\vec n_{j}=\vec p_{k}+\alpha_{k}(\vec p_f-\vec p_{k}),\quad \alpha_{k}=\frac{\delta}{|\vec p_f-\vec p_{k}|}.$$
Condition ensures that $\alpha_{k}\le1$ so that
$$\sum_{k}\frac{\vec p_f-\vec q_{k}}{|\vec p_f-\vec q_{k}|}
=\sum_{k}\frac{1-\alpha_{k}}{|1-\alpha_{k}|}\cdot \frac{\vec p_f-\vec p_{k}}{|\vec p_f-\vec p_{k}|}=\sum_{k}\frac{\vec p_f-\vec p_{k}}{|\vec p_f-\vec p_{k}|}=0$$
showing that $\vec q_f=\vec p_f$. As a result we have the joint measurement condition 
$$\sum_{k=0}^3|\vec q_{k}-\vec p_f|=\sum_{k=0}^3(1-\alpha_{k})|\vec p_{k}-\vec p_f|=4.$$ 
Furthermore, we actual have qubit observables, i.e., $|\vec n_k|\le 1$ which follows from joint measurement condition 
$$|\vec n_{j}|=\frac14\left|{\textstyle\sum_{k}}\gamma_{jk}(\vec q_{k}-\vec p_f)\right|\le
\frac14\sum_{k}|\vec q_{k}-\vec p_f|= 1$$ 
And the optimal lower bound can be attained in the pure state  with Bloch vector
$\vec r=\vec e_{0}$ where $$\vec e_{k}:=\frac{\vec p_f-\vec p_{k}}{|\vec p_f-\vec p_{k}|},\quad (k=0,1,2,3)$$
satisfying $\sum_k \vec e_k=0$ by the definition of the FT point. In fact we have the total uncertainty
\begin{eqnarray*}
\sum_{j=1}^3|\vec r\cdot (\vec m_{j}-\vec n_{j})|=
\frac\delta4\sum_{j=1}^3\left|\textstyle\vec e_{0}\cdot \sum_{k=0}^3\gamma_{jk}\vec e_{k}\right|\\
=
\frac\delta2\sum_{j=1}^3|\vec e_{0}\cdot (\vec e_{0}+\vec e_{j})|=
\frac\delta2\sum_{j=1}^3(1+\vec e_{0}\cdot \vec e_{j})=\delta\hskip-1.1cm.
\end{eqnarray*}
which means that the MUR is saturated. \hfill $\square$

As an optimal measurement always lies on the boundary, i.e., saturating the joint measurement condition, we may accomplish the optimal joint measurement in a single-qubit experiment~\cite{Yu2014arXiv}. 

{\bf Theorem 2} (Implementation) A jointly measurable triplet of unbiased qubit observables $\{ N_i\}$ that saturates the joint measurement condition Eq.(\ref{FMC}) can be implemented by the following parent measurement $\{P_kO_{k}\}$ with $O_k=\{O_{\mu_k|k}\}$ where
\begin{equation}\label{Implementation}
P_k=\frac{|\vec q_k-\vec q_f|}4,\quad O_{\mu_k|k}= \frac12\left(1+\mu_k\frac{\vec q_k-\vec q_f}{|\vec q_k-\vec q_f|}\cdot\vec\sigma \right), 
\end{equation}
with outcome labeled with $\mu_k=\pm1$ for each $k=0,1,2,3$ and post-measurement processing
$p_j(\mu|k,\mu_k)=\frac{1+\mu\ \gamma_{jk}\ \mu_k}2.$

{\bf Proof } The measurement can be implemented by first randomly choose $k=0,1,2,3$ according to distribution $P_k=\frac{|\vec q_k-\vec q_f|}4$ and then perform a measurement $\{ O_k=\{ O_{\pm|k}\}\}$ with 
$$ {{O}}_{\pm|k}=\frac{  {I}\pm\vec{g}_k\cdot\vec{\sigma }}{2},\quad \vec g_k=\frac{\vec q_k-\vec q_f}{|\vec q_k-\vec q_f|}.$$
In fact by using post-measurement processing
$$p_j(\mu|k,\mu_k)=\frac{1+\mu\ \gamma_{jk}\ \mu_k}2$$
we obtain the desired marginals
\begin{equation*}
	\sum_{k,\mu_k}p_j(\mu|k,\mu_k)   M(k,\mu_k)=\sum_kP_k\frac{1+\mu\gamma_{jk}\vec g_k\cdot\vec\sigma}2
	=   N_{\mu|j}
\end{equation*}
The standard parent measurement with 8 outcomes can be constructed as
\begin{eqnarray*}
   M(\omega_1,\omega_2,\omega_3)&=&\sum_{k,\mu_k}P_k\prod_{j=1}^3p_j(\omega_j|k,\mu_k)  {{O}}_{\mu_k|k}\\&=&\frac{1+\sum_j\omega_j\vec n_j\cdot\vec\sigma-\omega_1\omega_2\omega_3\vec q_f\cdot\vec\sigma}8.
\end{eqnarray*}
with $\omega_j=\pm$, which coincides with the joint measurement given in Ref.~\cite{Yu2013arXiv}.\hfill $\square$

As an immediate result, when the MUR is attained the optimal joint measurement Eq.(\ref{optm}),  can be readily implemented by performing 4 ideal qubit measurements along directions $\vec e_k$ with probability $\frac14(|\vec p_k-\vec p_f|-\delta)$ for $k=0,1,2,3$.

\section{Symmetry}
Symmetry is considered to be the main characteristics of the laws of physics according to Feynman~\cite{Feynman} and plays an essential role in determining incompatibility noise robustness~\cite{Toigo2018,Guhne2020}.
Some physically relevant sets of measurements may admit certain symmetries among measurement directions.  For example in the case of an ideal triplet $\mM_\perp$ as shown in Fig.1 with one observable ($\vec m_3$ along $\vec{x}$ direction) being orthogonal to the other two ($\vec m_{1,2}$ on the $YZ$ plane) the reflection $\tau_{YZ}$ over $YZ$ plane and  the reflections $\tau_{\pm}$ over planes passing through $\vec z$ and angle bisectors $\vec m_1\pm\vec m_2$ of $\vec m_{1,2}$ are a kind of graded symmetry. By a graded symmetry here we mean a reflection such that three directions are permuted among themselves up to some inversions (which correspond to relabeling two outcomes of a measurement). For example, we have
$\tau_{YZ}\cdot \vec m_3=-\vec m_3$ while preserving directions $\vec m_{1,2}$ and $\tau_\pm\cdot\vec m_{k}=\pm\vec m_{3-k}$ with $k=1,2$ while preserving $\vec m_3$.  It turns out that the optimal measurement must share the same graded symmetry.
\begin{figure}
\includegraphics[width=0.45\textwidth]{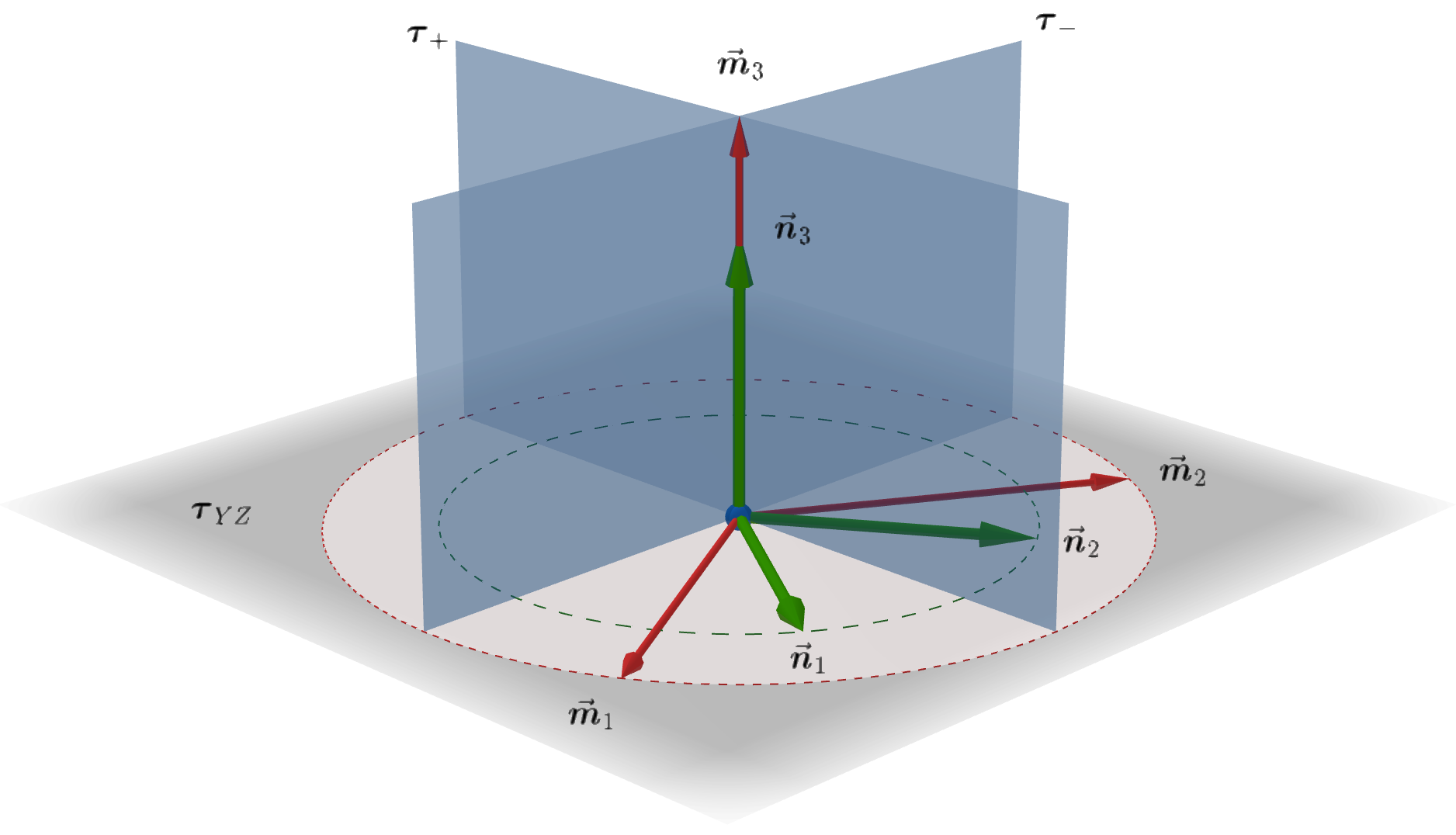}
\caption{Target triplet $\mM_\perp$ together with its symmetries and the optimal triplet $\mN_\perp$ with the same symmetries.}
\end{figure}

{\bf Theorem 3} (Symmetry) If the ideal triplet $\mM$ along directions $\{\vec m_j\}_{j=1}^3$ admits a graded symmetry $g$, i.e., a reflection such that
$g\cdot \vec m_j={\omega_j} \vec m_{\sigma(j)}$
with  $\omega_j=\pm$ and $\sigma$ being a permutation of $\{1,2,3\}$, then the optimal jointly measurable triplet $\mN=\{\vec n_j\}_{j=1}^3$ for the incompatibility $\Delta_\mM$ shares the same graded symmetry, i.e.,
$g\cdot \vec n_j={\omega_j}  \vec n_{\sigma(j)}.$

{\bf Proof } Firstly, we note that as $g$ is a reflection we have $g^2=1$ so that $\vec m_j=g\cdot g\cdot \vec m_j=g\cdot \omega_j\vec m_{\sigma(j)}=\omega_j\omega_{\sigma(j)}\vec m_{\sigma^2(j)}$ from which it follows that $\sigma^2(j)=j$ and $\omega_j=\omega_{\sigma(j)}$. 
Under the graded symmetry $g$ we have
\begin{eqnarray*}
g\cdot \vec m_\mu&=&g\cdot \sum_{j=1}^3\mu_j\vec m_j=\sum_{j=1}^3\mu_jg\cdot\vec m_{j}\\
&=&\sum_{j=1}^3\mu_j\omega_j\vec m_{\sigma(j)}=\sum_{j=1}^3\mu_{\sigma(j)}\omega_{\sigma(j)}\vec m_{j}=\vec m_{\mu_\sigma \omega_\sigma}.
\end{eqnarray*}
In order to show that the optimal joint measurement also has the same symmetry it suffices to show for each joint measurement  there exists another joint measurement having the symmetry and in the mean time with a no larger uncertainty. In fact, if $\{\vec n_j\}$ is an arbitrary jointly measurable triplet of unbiased observables, then we have
\begin{eqnarray}
&&2\max_{\mu}|\vec m_{\mu}-\vec n_{\mu}|\nonumber\\&=&\max_\mu|\vec m_{\mu}-\vec n_{\mu}|+\max_\mu|g\cdot(\vec m_{\mu}-\vec n_{\mu})|\nonumber\\
&=& \max_\mu|\vec m_{\mu}-\vec n_{\mu}|+\max_{\mu}|\vec m_{\mu_\sigma\omega_\sigma}-g\cdot \vec n_{\mu}|\nonumber\\
&=& \max_\mu|\vec m_\mu-\vec n_{\mu}|+\max_{\mu}|\vec m_{\mu}-g\cdot \vec n_{\mu_\sigma\omega}|\nonumber\\
&\ge&2\max_\mu\left|\vec m_\mu-\frac{\vec n_\mu+g\cdot \vec n_{\mu_\sigma\omega}}2\right|.\label{in}
\end{eqnarray}
Therefore if we consider another triplet of unbiased observables along directions
$$\vec n'_{j}=\frac{\vec n_j+\omega_j g\cdot \vec n_{\sigma(j)}}2$$
we have $g\cdot \vec n'_j=\omega_j\vec n'_{\sigma(j)}$ (by noting $\sigma^2(j)=j$ and $\omega_j=\omega_{\sigma(j)}$), i.e., the averaged triplet $\{\vec n'_j\}$ shares the same symmetry with triplet $\mM$. In order to  show that $\{\vec n'_j\}$ is also jointly measurable we note that the joint measurement condition for $\{\vec n_j\}$ can be equivalently written as
\begin{equation}\label{ftsy}
\sum_{\mu}|\vec n_\mu-\mu_0\vec q_f|\le8\Leftrightarrow \sum_{\mu}|g\cdot\vec n_{\mu_\sigma\omega}-\omega_0\mu_0g\cdot \vec q_f|\le8,
\end{equation}
where we have denoted $\mu=(\mu_1,\mu_2,\mu_3)$ and $\mu_0=\mu_1\mu_2\mu_3$ and $\omega_0=\omega_1\omega_2\omega_3$.
By summing up these two inequalities and using triangle inequality we have 
$$8\ge\sum_{\mu}\left|\vec n_\mu'-\mu_0\frac{\vec q_f+\omega_0g\cdot\vec q_f}2\right|\ge\sum_{\mu}\left|\vec n_\mu'-\mu_0\vec q'_f\right|,$$
where
$$ \vec n'_\mu=\sum_j\mu_j\vec n_j'=\frac{\vec n_\mu+g\cdot \vec n_{\mu_\sigma\omega}}2$$ 
with $\vec q_f'$ being the FT point of $\{\vec q'_k\}=\{\vec n'_\mu|\mu_0=1\}$,
which gives the desired joint measurability of $\{\vec n'_j\}$. In sum, starting form an arbitrary jointly measurable triplet $\{\vec n_j\}$  the newly introduced triplet $\{\vec n_j'\}$ is also jointly measurable, shares the same symmetry as $\mM$, and, due to Eq.(\ref{in}), has a  no larger total uncertainty. Therefore the optimal measurement can be taken to share the same symmetry as $\mM$ without loss of generality. \hfill$\square$

We note that Eq.(\ref{ftsy}) actually proves that if the jointly measurable triplet $\{\vec n_j\}$ possesses some graded symmetry $g$, i.e., $g\cdot \vec n_j=\omega_j\vec n_{\sigma(j)}$, then the FT point also shares the same graded symmetry in the sense that $g\cdot\vec p_f=\omega_0\vec p_f$.

\section{Two analytical examples}

Some triplets of unbiased observables might be determined completely by their symmetry. For example the triplet of three orthogonal observables, e.g., $\{\sigma_k\}$, is completely determined by 3 reflections over planes $XY,YZ,ZX$ (upto some scalings). In this case, the symmetry theorem above therefore enables us to determine the optimal measurements completely (upto some scalings). In other cases however the symmetry might determine partially the triplet. And in this case the symmetry also help to simplify the optimization. Our first example is the triplet $\mM_\perp$ along directions
$$\vec m_1=(\sin   \theta,0,\cos   \theta),\quad \vec m_2=(1,0,0),\quad \vec m_3=(0,1,0)$$

The reflection $\tau_{YZ}$ over $YZ$ plane and  the reflections $\tau_{\pm}$ over planes passing $\vec x$ and angle bisectors $\vec m_\pm=\vec m_1\pm\vec m_2$ of $\vec m_{1,2}$ generate the graded symmetry group of this triplet. In fact, we have
$\tau_{YZ}\cdot \vec m_3=-\vec m_3$ while preserving directions $\vec m_{1,2}$ and $\tau_\pm\cdot\vec m_{k}=\pm\vec m_{3-k}$ with $k=1,2$ while preserving $\vec m_3$. Using theorem above we see that the optimal joint measurement $\{\vec n_j\}$ should also have such a symmetry. From $\tau_{XY}\cdot \vec n_3=-\vec n_3$ it follows that $\vec n_3$ should be also along the direction given by the eigenvector of $\tau_{YZ}$ corresponding eigenvalue -1, which infers $\vec n_3\propto \vec m_3$. Similarly from $\tau_{YZ}\cdot \vec n_k=\vec n_k$ it follows that $\vec n_{1,2}$ also lie on the $YZ$ plane. Finally, from symmetry $\tau_\pm\cdot\vec n_{k}=\pm\vec n_{3-k}$ we see that $\vec n_{1,2}$ have the same length and are  also symmetric about the angle bisect $\vec m_\pm$. As a result the optimal joint measurement must be of form
$$\vec n_3=n_3\hat m_3,\ \vec n_1=\frac{\beta_+\hat m_++\beta_-\hat m_-}2,\ \vec n_2=\frac{\beta_+\hat m_+-\beta_-\hat m_-}2,$$
where $$\hat m_\pm=\frac{\vec m_1\pm \vec m_2}{|\vec m_1\pm \vec m_2|},\quad \hat m_3=\frac{\vec m_3}{|\vec m_3|}$$
with 3 suitable constants $n_3,\beta_\pm$ satisfying joint measurement condition
$|\beta_+|+|\beta_-|\le2\sqrt{1-n_3^2}.$ 
Now the incompatibility $\Delta_{\perp}$ can be calculated as
\begin{eqnarray*}
&&\min_{n_3,\beta_\pm} \max_{\pm}\sqrt{(1-n_3)^2+(\beta_\pm-m_\pm)^2}\\
&=&\min_{n_3,\beta_\pm} \max_{\pm}\sqrt{(1-n_3)^2+(|\beta_\pm|-m_\pm)^2}\\
&=&\min_{n_3,\kappa} \sqrt{(1-n_3)^2+(d_+-\sqrt{1-n_3^2}+ |\kappa|)^2}\\
&=&\left\{\begin{array}{ll}\displaystyle \min_{n_3}\sqrt{(1-n_3)^2+(d_+-\sqrt{1-n_3^2})^2} &  d_-\le\sqrt{1-n_3^2}\\
\displaystyle \min_{n_3}\sqrt{(1-n_3)^2+(m_+-2\sqrt{1-n_3^2})^2}& d_->\sqrt{1-n_3^2}\end{array}\right.
\end{eqnarray*}
In the above we have denoted $$m_\pm=|\vec m_\pm|,\quad d_\pm=\sqrt{1\pm\cos   \theta}$$ so that the joint measurement condition gives
$$|\beta_\pm|=\sqrt{1-n_3^2}\pm ( d_--\kappa)$$
for some real $\kappa$ satisfying
$$ d_--\sqrt{1-n_3^2}\le\kappa\le d_-+\sqrt{1-n_3^2}.$$
In the first case $ d_-\le\sqrt{1-n_3^2}$ we have to minimize
\begin{eqnarray*}
 \sqrt{(1-n_3)^2+(d_+-\sqrt{1-n_3^2})^2}\hskip 1.1cm\\
 =\sqrt{2+d_+^2-2(n_3+d_+\sqrt{1-n_3^2})}\hskip -1.1cm
 \end{eqnarray*}
on the condition $ d_-\le \sqrt{1-n_3^2}$. By Cauchy-Schwarz inequality we have minimal value $\sqrt{1+d^2_+}-1$ as long as the optimal $n_3^*=\frac1{\sqrt{1+d^2_+}}$ satisfies the constraint which reads 
$1\le (1- d_-^2)(1+d^2_+)$
giving rise to condition $\cos   \theta\ge\sqrt2-1$, i.e., $   \theta\le    \theta_0\approx 65.53^\circ$.
Beyond this range, CS inequality cannot be attained so that the minimal value is taken by boundary value $ d_-=\sqrt{1-n_3^2}$ which gives the second bound
\begin{eqnarray*}
\sqrt{(1-\sqrt{1- d_-^2})^2+(d_+- d_-)^2}
\hskip 1.1cm\\=\sqrt{3+\cos   \theta-2\sqrt{\cos   \theta}-2\sin   \theta}\hskip-1.1cm
\end{eqnarray*}
While the first case gives the first two bounds, the second case gives the third bound
Let $n_3=\cos t$ for the minimization of the second case we have (by derivative)
$$\sin   \theta=\frac18\tan^2t(1+3\cos t)^2-1\ge \sqrt{1-\cos^4t} $$
with equality holding for $   \theta=   \theta_1\approx 71.53^\circ$. The third bound therefore can be implicitly given by a parametric curve $\Delta_{opt}(   \theta)$
\begin{eqnarray*}\Delta_{opt}&=&\frac{1-\cos t}{\cos t}\sqrt{1+3\cos^2t}\\
   \theta&=&\arcsin\left(\frac18\tan^2t(1+3\cos t)^2-1\right)
\end{eqnarray*}
To summarize the compatibility reads
\begin{equation}\label{sym1}
\Delta_{\perp}=\left\{\begin{array}{ll}{2\sqrt{2+\cos   \theta}-2} &    \theta\le   \theta_0\\
{2\sqrt{3+\cos   \theta-2\sqrt{\cos   \theta}-2\sin   \theta}}&    \theta_0\le   \theta\le    \theta_1\\
{\Delta_{opt}(   \theta)}&   \theta\ge    \theta_1\end{array}\right..
\end{equation}
Respectively, there are three optimal measurements for three ranges  of $   \theta$:
\begin{itemize}
\item[M1] $   \theta\le   \theta_0$. In this case the condition of Theorem 2 is met and the MUR can be saturated so that the optimal measurement, as given by Theorem 2, reads
$$\vec n_3=\frac{\vec m_3}{\sqrt{d^2_++1}},\ \vec n_{1,2}=\frac{\beta_+^*\hat m_+\pm\beta_-^*\hat m_-}2,$$
where $ \beta_\pm^*=\frac{d_+}{\sqrt{d^2_++1}}\pm d_-. $
\item[M2] $   \theta_1\ge   \theta\ge   \theta_0$. In this case MUR cannot be saturated and the optimal measurement reads
$$\vec n_3=\sqrt{1-d_-^2}\ \vec m_3,\quad \vec n_1=\vec n_2=d_- \ \hat m_+$$
\item[M3] $   \theta\ge   \theta_1$. The optimal joint measurement reads
$$\vec n_3=\sqrt{1-(d_--\kappa)^2}\ \vec m_3,\quad \vec n_1=
 \vec n_2=(d_--\kappa)\hat m_+$$
 with $\kappa\ge0$ minimizing $$\sqrt{(1-\sqrt{1-(d_--\kappa)^2})^2+(m_-+2\kappa)^2}.$$
\end{itemize}
If we introduce $\cos2\gamma=\vec m_1\cdot\vec m_2=\sin\theta$ and three corresponding regions for different optimal measurements are determined by which one of the following three intervals $$0<\Gamma_0<\Gamma_1<90^\circ$$ that $|\gamma-45^\circ|$ falls into, 
where
$$\Gamma_j=45^\circ-\frac{180^\circ}{2\pi}{\arcsin\cos\theta_j},\quad (j=0,1).
$$
\begin{figure}
\includegraphics[width=0.41\textwidth]{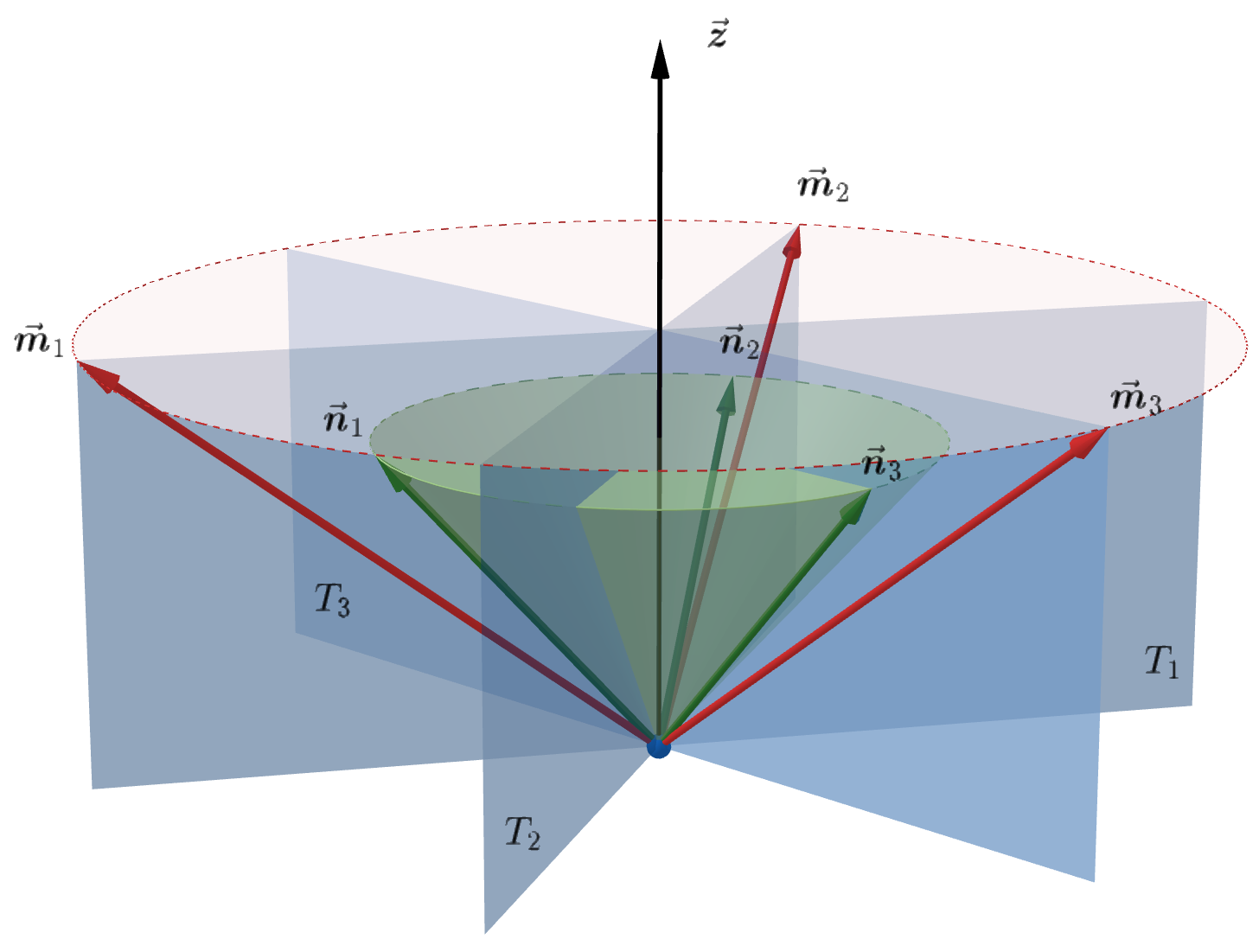}
\caption{Target triplet $\mM_Y$ together with its symmetries and the optimal triplet $\mN_Y$ with the same symmetries.}
\end{figure}
Our second example is  the following highly symmetric triplet $\mM_Y$ of unbiased observables along directions
$$\vec m_j=\hat z\cos\gamma+\hat e_j\sin\gamma,\quad \hat e_j\cdot\hat e_k=-\frac 12,\quad \hat z\cdot\hat e_j=0,$$
with $0\le\gamma\le\frac\pi2$ and $\hat z$ is the unit vector along $Z$ direction, as illustrated in Fig.2.
Firstly, let us see under what conditions the MUR $\Delta_Y\ge 2\delta_\perp$ for this triplet can be saturated. Four FT anchors are
$\vec p_0=3\cos\gamma\hat z$ and $ \vec p_j=-\cos\gamma\hat z+2\sin\gamma\vec e_j$ for $j=1,2,3$ with
 FT point reads $\vec p_f=\alpha \hat z$ (by symmetry) where
$$\alpha=\left\{\begin{array}{lcl} \frac {\sin\gamma}{\sqrt2}-\cos\gamma &\quad& 0\le\tan\gamma\le 4\sqrt 2\\
3\cos\gamma &\quad&|\tan\gamma|\ge 4\sqrt2\\
-\frac {\sin\gamma}{\sqrt2}-\cos\gamma &\quad& 0\ge\tan\gamma\ge -4\sqrt2\end{array}\right.$$
(a detailed derivation given later) giving rise to the lower bound
$\Delta_\perp\ge2\delta_\perp$ where $$\delta_\perp=\left\{\begin{array}{cc}|\cos\gamma|+\sqrt 2\sin\gamma-1 & |\tan\gamma|\le4\sqrt2\\
\frac32\sqrt{4\cos^2\gamma+\sin^2\gamma}-1
&|\tan\gamma|\ge 4\sqrt2\end{array}\right. $$
In the case of $\gamma\le\gamma_0=\arctan 2\sqrt2\approx 70.53^\circ$, which is determined by the condition $\delta=\min_k|\vec p_k-\vec p_f|$, namely, $\sin\gamma/\sqrt2-\cos\gamma=1/3$, the attainability condition of Theorem 1 is satisfied so that the MUR can be attained. 

In the case of $\gamma>\gamma_0$ MUR cannot be attained. In order to find the optimal measurement for the incompatibility we consider the symmetry of triplet $\mM_Y$. Let  $T_j$ be the reflection over the plane orthogonal to $\hat z\times \hat e_j$ passing the origin. Then we have $T_j\cdot \vec m_j=\vec m_j$ while $T_j\cdot \vec m_k=\vec m_l$ with $(j,k,l)$ cyclic. This symmetry is obeyed by the optimal joint measurement triplet according to Theorem 3. As a result the optimal triplet $\mN_Y$  should be along directions 
\begin{equation}\label{opt2}
\vec n_j=x\hat z+y\hat e_j,\quad j=1,2,3.
\end{equation}
Next we shall prove that the triplet above is jointly measurable if and only if 
\begin{equation}\label{xy} y\le \left\{\begin{array}{ll} \frac{1-x}{\sqrt2}& |x|\ge \frac19\\
\frac23\sqrt{1-9x^2}& |x|\le \frac19\end{array}\right.
\end{equation}
In fact four FT anchors for this triplet are
$\vec q_0=3x\hat z,\ \vec q_j=-x\hat z+2y\hat e_j$
with FT point being $\vec q_f=\alpha \hat z$ (by symmetry) so that we have total distance
$$\sum_k|\vec q_k-\vec q_f|=|3x-\alpha|+3\sqrt{(x+\alpha)^2+4y^2}$$
which is minimized by (via derivative as a function of $\alpha$)
$$\alpha=\left\{\begin{array}{ll} \frac y{\sqrt2}-x & 4\sqrt2 x\ge y\\
3x &4\sqrt 2|x|\le y\\
-\frac y{\sqrt2}-x & 4\sqrt 2x\le -y\end{array}\right.$$
giving rise to the joint measurement condition
$$4\ge\sum_{k=0}^3|\vec q_k-\vec q_f|=\left\{\begin{array}{cc} 4|x|+4\sqrt2 y & 4\sqrt 2|x|\ge y\\
6\sqrt{4x^2+y^2} & 4\sqrt 2|x|\le y\end{array}\right.,$$
which leads to the desired condition Eq.(\ref{xy}) for $x,y$.

According to Theorem 3 the optimal incompatibility can be obtained by considering only triplet of from Eq.(\ref{opt2}), which now becomes
$$\frac{\Delta_Y}2=\min_{x,y}\max\{3|\cos\gamma-x|,\sqrt{(\cos\gamma-x)^2+4(\sin\gamma-y)^2}\}$$
with minimum taken over region determined by Eq.(\ref{xy}). By min-max inequality we have 
$$\frac{\Delta_Y}2\ge \min_{x,y}\sqrt{(\cos\gamma-x)^2+4(\sin\gamma-y)^2}$$
whose minimum is attained at (by derivative over the boundary $y=\frac23\sqrt{1-9x^2}$) 
\begin{equation}\label{der}
(\cos\gamma-x)y=16(\sin\gamma-y)x.
\end{equation} Furthermore in order for $\Delta_Y$ to attain this lower bound we must also have
\begin{equation}\label{minmax}
3|\cos\gamma-x|\le\sqrt{(\cos\gamma-x)^2+4(\sin\gamma-y)^2}
\end{equation}
which is possible as long as $\gamma\ge\gamma_1$ where
$$\gamma_1=\arctan\frac{1}{79} \sqrt{2} \left(5 \sqrt{337}+129\right)\approx 75.80^\circ$$
is determined by Eq.(\ref{der}) joint measurability condition $y=\frac23\sqrt{1-9x^2}$ together with the equality in Eq.(\ref{minmax}).
 
\begin{figure}
\includegraphics[width=0.45\textwidth]{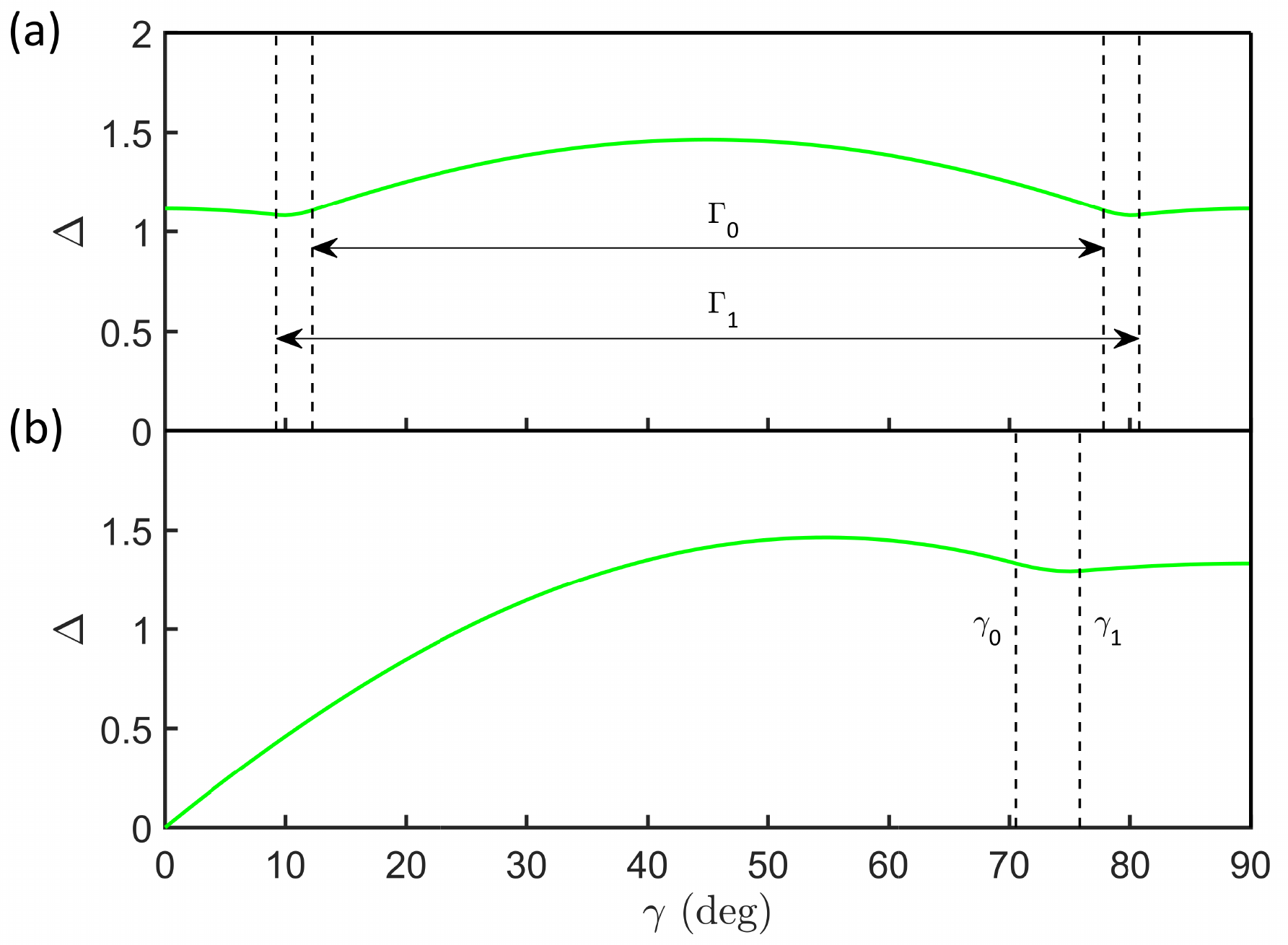}
\caption{Exact values of incompatibility for two triplets of ideal qubit measurements with symmetries, (a) triplet $\mM_\perp$ with $|\cos2\gamma|=\vec m_1\cdot\vec m_2$ and (b) highly symmetric triplet $\mM_Y$ of unbiased observables (see main text for details) according to analytical results computed with Eqs. (\ref{sym1}) and (\ref{sym2}), respectively. }
	\label{fig:F-sym}
\end{figure} 
 
Though not all conditions in Eq.(\ref{unn}) can be satisfied simultaneously, some of them, e.g., condition 1, can be satisfied, which leads to 
$${\Delta_Y}\ge \min_{x,y}3|\cos\gamma-x|+\sqrt{(\cos\gamma-x)^2+4(\sin\gamma-y)^2}$$
which is attained by $(x,y)$ equalizing Eq.(\ref{minmax}) and  $y=\frac23\sqrt{1-9x^2}$, giving rise to
$$\frac{\Delta_Y}2=\frac{\sin \gamma}{\sqrt2}+2 | \cos\gamma|- \sqrt{\frac{2}{3}-\big(\sin\gamma-\sqrt{2} | \cos\gamma| \big)^2}$$
This happens if $\gamma_0\le\gamma\le\gamma_1$. 
To summarize, the incompatibility reads
\begin{equation}\label{sym2}
	\Delta_{Y}=\left\{\begin{array}{ll}{2\cos\gamma+2\sqrt 2\sin\gamma-2} \\
		{\sqrt2{\sin \gamma}+4\cos\gamma- 2\sqrt{\frac{2}{3}-\left(\sin\gamma-\sqrt{2} \cos\gamma \right)^2}}\\
		{\displaystyle\min_{x^2+y^2=\frac19}2\sqrt{(\cos\gamma-x)^2+4(\sin\gamma-2y)^2}}
	\end{array}\right.
\end{equation}
for three intervals $0<\gamma_0<\gamma_1<90^\circ$ divided by $\gamma_0\approx 70.53^\circ$ and $\gamma_1\approx75.80^\circ$.
In Fig.3 we plot the exact incompatibility measures of these two symmetric triplets, which fit perfect with the numerical results.

\section{Conclusions an discussions} Quantum incompatibility reflects the basic fact that 
some quantum measurements may disturb each other, preventing us from measuring them  with a single measurement device without introducing errors. The MUR sets the limit to how well we can perform the joint measurement with the minimal amount of errors according to quantum mechanics, providing a quantitative characterization of quantum incompatibility. 
In establishing rigorously the triplet MUR we also find out the necessary and sufficient conditions under which the MUR is attained and, when the attainability condition is met, we also explicitly construct the optimal jointly measurable triplet. In order to facilitate experimental tests we design a straightforward implementation of optimal joint measurements by just performing randomly four ideal qubit measurements. Lastly, symmetry is found to play an essential role in deriving analytically the incompatibility measure.

 As a critical step in the study of joint-measurement on multiple observables, this study may deepen our understanding of Heisenberg's measurement uncertainty principle and lead to new applications in quantum metrology and quantum information science. We stress that the demonstrated strategy can be readily generalized to calculate the incompatibility of four or more qubit observables and the case of weighted distance as uncertainty measure, which will be considered elsewhere. 

\section*{Acknowledgement}
This work is supported by the Key-Area Research and Development Program of Guangdong Province Grant No.2020B0303010001, Grant No.2019ZT08X324, National Natural Science Foundation of China Grants No.12004207 and No.12005090, Shenzhen Science and Technology Program Grant No.RCYX20210706092043065, Guangdong Provincial Key Laboratory Grant No.2019B121203002 and SIQSE202104. 

\newpage

\bibliography{Ref-3O}

\end{document}